\documentclass[pra, preprint, superscriptaddress]{revtex4}
\usepackage[latin2]{inputenc}
\usepackage[letterpaper]{geometry}
\usepackage[dvipsnames]{xcolor}
\usepackage{amsmath,amssymb,amsfonts,mathtools,accents}
\usepackage{graphicx}
\usepackage{setspace}
\usepackage{verbatim}
\usepackage[normalem]{ulem}
\usepackage{cancel}
\usepackage{bm}

\usepackage{tikz} 
\tikzset{every picture/.style={remember picture}}
\usetikzlibrary{positioning, backgrounds, calc, arrows.meta}
\usepackage{tikzsymbols}  
\usepackage{tikzorbital}

\usepackage[breaklinks=true,hidelinks]{hyperref}
\hypersetup{
  colorlinks=true,
  linkcolor=blue,
  filecolor=magenta,
  urlcolor=cyan,
  allcolors=blue
}

\definecolor{MyColor}{rgb}{0,0.7,0}

\begin{document}

\title{Light-induced nonadiabatic photodissociation of the 
$\mathrm{NaH}$ molecule including electron-rotation coupling}

\author{Zolt\'{a}n Kir\'{a}ly}
\affiliation{Department of Theoretical Physics, Doctoral School of Physics, University of Debrecen, H-4002 Debrecen, PO Box 400, Hungary}

\author{Otabek Umarov}
\affiliation{Department of Exact Sciences, Kimyo International University in Tashkent, Shota Rustaveli Street 156, Tashkent 100121, Uzbekistan}

\author{Csaba F\'{a}bri}
\affiliation{Department of Theoretical Physics, Doctoral School of Physics, University of Debrecen, H-4002 Debrecen, PO Box 400, Hungary}

\author{G\'{a}bor J. Hal\'{a}sz}
\affiliation{Department of Information Technology, University of Debrecen, H-4002 Debrecen, PO Box 400 Hungary}

\author{Attila T\'{o}th}
\affiliation{ELI ALPS, The Extreme Light Infrastructure ERIC, Wolfgang Sandner str. 3, 6728 Szeged, Hungary}

\author{\'{A}gnes Vib\'{o}k}
\affiliation{Department of Theoretical Physics, Doctoral School of Physics, University of Debrecen, H-4002 Debrecen, PO Box 400, Hungary}
\affiliation{ELI ALPS, The Extreme Light Infrastructure ERIC, Wolfgang Sandner str. 3, 6728 Szeged, Hungary}
\email{vibok.agnes@science.unideb.hu}

\begin{abstract}
It is well established that electronic conical intersections (CIs)
in molecular systems can be induced by laser light, even in diatomic
molecules. The emergence of these light-induced degeneracies leads
to strong coupling among electronic, vibrational, and photonic modes,
which significantly influences ultrafast nuclear dynamics. In this
work, we perform pump--probe numerical simulations on the NaH molecule,
considering the first three singlet electronic states--- ($\mathrm{X^{1}\mathrm{\Sigma}^{+}(X)},$
$\mathrm{A^{1}\mathrm{\Sigma}^{+}(A)}$ and $\mathrm{B^{1}\Pi(B)}$)
---and including several light-induced degeneracies in the theoretical
model. To elucidate the ultrafast molecular dynamics, the combined
effects of multiple light-induced nonadiabatic couplings and rotational
motion of the nuclei together with the situation when the electronic
angular momentum projected onto the diatomic axis couples with the
angular momentum of the nuclei has been studied. We then calculate 
key dynamical observables such as dissociation probabilities, 
kinetic energy release spectra, and angular distributions of the 
photofragments within and above the linear regime.

\end{abstract}

\maketitle

\section{Introduction}

Understanding the behavior of atoms and molecules under strong electromagnetic
fields has been an extensively investigated area of research. A large
number of theoretical and experimental studies have been devoted to
exploring a variety of novel phenomena arising from light--matter
interactions. Many of these works focus on the dynamical behavior
of diatomic systems, beginning with the simplest hydrogen-like ions
or molecules and extending to systems containing a large number of
electrons \cite{Bandrauk1,Bandrauk2,Takasuka1,Takasuka2,Tiwari0,Takasuka3,Tiwari1,Attila1,
Zhaopeng1,Foudil1,Attila2,Attila3,Zhaopeng2,Zhang1,Zhang2,Zhang3,Zhang4,Umarov1}.
Nevertheless, numerous other important studies have also addressed
the photodissociation and fragmentation of polyatomic molecules 
\cite{Banares4,Banares5,Fabien2,Graham2,Ignacio2,Ignacio3,Weinacht1,Weinacht2,Weinacht3,Yarkony1}.

Molecular dissociation is commonly analyzed within the Born--Oppenheimer
(BO) framework, which is based on the separation of electronic and
nuclear motions due to their distinct time scales. In most cases,
this approximation provides a satisfactory description of dynamical
processes. However, at certain nuclear configurations---particularly
in the vicinity of degeneracy points or conical intersections (CIs)---the
energy exchange between electrons and nuclei becomes significant,
leading to a breakdown of the BO approximation \cite{Born1}. Conical
intersections play a crucial role in nonadiabatic processes, serving
as efficient funnels for ultrafast interstate transitions that typically
occur on the femtosecond timescale \cite{Koppel1,Yarkony2,Graham1,Koppel2,Baer1,Morgane1}.

CIs in molecules can only form when a molecular system possesses at
least two independent nuclear degrees of freedom. Consequently, since
a diatomic molecule has only a single vibrational coordinate, CIs
cannot arise in such systems under field-free conditions. However,
this holds only in free space. When a diatomic molecule is exposed
to a strong laser field, its electronic energy becomes dependent on
the angle between the laser polarization direction and the molecular
axis. This occurs because the interaction of the induced or permanent
dipole moment with the external field generates an effective torque
that tends to align the molecule with the field polarization.

Previous studies have demonstrated that laser waves can induce CIs
even in diatomic molecules. In these cases, the rotational motion
provides the additional degree of freedom required for the formation
of a light-induced conical intersection (LICI) \cite{Nimrod1}. The
presence of a LICI introduces strong nonadiabatic couplings through
the mixing of rotational, vibrational, electronic and photonics modes.
Importantly, the energetic and spatial locations of such LICIs can
be controlled by tuning the laser field parameters, such as its intensity
and frequency. Depending on the level of theoretical description,
whether the rotational degree of freedom is accounted for as a parameter
or a dynamic variable, the interaction with light can manifest as
light-induced avoided crossing (LIAC) or LICI. In a restricted one-dimensional
(1D) description, the molecular rotational angle is only a parameter
and thus can only mimic a light-induced avoided crossing situation
(LIAC). However, in the full two-dimensional (2D) description, the
rotational angle is a dynamic variable, the inclusion of which leads
to a LICI explicitly included in the description. Numerous theoretical
and experimental results in the past few years undoubtedly demonstrated
that LICIs exert strong effects on the spectroscopic, quantum dynamical
and topological properties of both diatomic \cite{Gabor1,Gabor2,Gabor3,Gabor4,
Andris1,Andris2,Andris3,Andris4,Tamas1,Bandrauk3,Buksbaum1,Nature1,Ding1,
Kirrander1,Zhaopeng3,Han1,Kotochigova1,Han2,Bandrauk5,Thumm1,Thumm2,Markus1}
and polyatomic systems \cite{Buksbaum2,Banares,Csaba1,Batista1,Tam=0000E1s3},
even for weak laser fields.

Recent publications \cite{Zhang1,Zhang2,Zhang3,Zhang4,Zhang5,Zhang6} 
pointed out that the coupling between the molecular rotational angular 
momentum ($\mathbf{K}$) and the total electronic angular momentum projected 
onto the molecular axis $\Omega=\Lambda+\Sigma$, i.e. $\mathbf{K}$-$\Omega$ 
coupling, might become important for the correct interpretation of 
rotation-related physical quantities. This effect is routinely neglected 
due to the large mass difference between molecules and electrons. 
However, at high laser intensities it can significantly alter the 
rotational degrees of freedom of molecules through repeated Rabi 
oscillations between the electronic states.

The goals of the present work are two-fold. First, efforts are made
to find the appropriate pump and probe laser parameters (frequency,
intensity, laser pulse length) that 
fall within or slightly above the linear regime. This makes analysis 
of the results easier.
Second, we perform
pump and probe numerical simulations to investigate the light-induced
nonadiabatic photodissociation dynamics of the NaH molecule \cite{NaH1,NaH2,NaH3,NaH4,Umarov1}
in the presence of nuclear as well as electron rotation. Next, in
order to properly discuss the dissociation dynamics, the dissociation
probabilities, kinetic energy release (KER) spectra, and photofragment 
angular distributions (PAD) are calculated. We perform 1D and 2D calculations
to model the LIAC and LICI, respectively, as well as so-called 3D
simulations, where beyond the nuclear rotational angular momentum 
we also incorporate the electronic interior angular momentum about 
the molecular-fixed $z$ axis in our description.

The article is organized as follows. In the next section, we describe
the methodology and algorithms employed in the theoretical analysis,
including the working Hamiltonian, the applied electric fields, the
computed dynamical quantities, and relevant numerical details. Section
III presents and discusses the numerical results obtained for the
NaH system. Finally, the main conclusions are summarized in the last
section.
Atomic units ($\mathrm{e=m_{e}=\hbar=1)}$ are used throughout the paper, 
unless stated otherwise.

\section{Computational details}

\begin{figure}[!ht]
\begin{centering}
\includegraphics[width=0.48\textwidth]{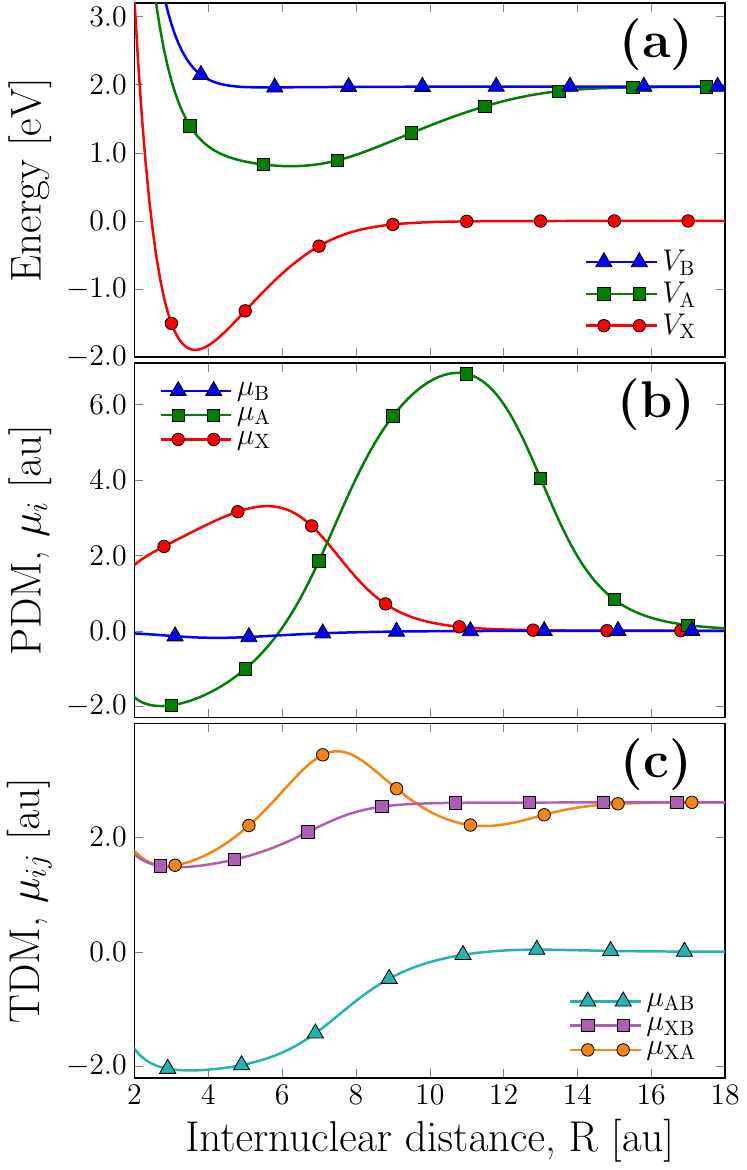}
\par\end{centering}
\caption{ (a) Lowest three adiabatic potential energy curves of the 
$\mathrm{NaH}$ molecule. (b) Permanent dipole moment (PDM) functions 
of the three adiabatic electronic states. (c) Transition dipole moment 
(TDM) functions between the different adiabatic electronic states.}\label{fig:fig1}
\end{figure}

Our showcase example is the NaH molecule. As in case of our previous work 
\cite{Umarov1} on the same system, our model includes the three lowest-lying 
singlet electronic states $\mathrm{X}^{1}\mathrm{\Sigma}^{+}$, 
$\mathrm{A^{1}\mathrm{\Sigma}^{+}}$ and $\mathrm{B^{1}\Pi}$, labeled throughout 
the paper as $\mathrm{X}$, $\mathrm{A}$ and $\mathrm{B}$, respectively.
$^{1}\Pi$ is a degenerate state. In the appendix we describe how it is used 
effectively as a single state in the calculations (Eq. \ref{Eq:tdm_final}).
Their corresponding potential energy curves $V_{\mathrm{X/A/B}}(R)$ are presented 
in Fig. \ref{fig:fig1} (a) along the 
$\vec{\mu}_{ii}(R) = \left\langle \phi_{i} \mid \Sigma_{k} \vec{r}_{k} \mid \phi_{i}\right\rangle$ 
permanent dipole moments (PDMs) on panel (b), and the 
$\vec{\mu}_{ij}(R)=\left\langle \phi_{i} \mid \Sigma_{k} \vec{r}_{k} \mid \phi_{j}\right\rangle $ 
transition dipole moments (TDMs) on panel (c), for 
$i,j\in\left\{\mathrm{X},\mathrm{A},\mathrm{B}\right\}$.
We note here that all permanent dipole moments of $\Sigma$ states and 
$\Sigma-\Sigma$ TDMs are parallel, whereas those that involve the $\Pi$ 
state are perpendicular to the molecular axis. Computation of the above 
electronic structure quantities of NaH has been carried out with the 
MOLPRO \cite{Molpro} program package at the MRCI/CAS(2/9)/aug-cc-pV5Z 
level of theory. The number of active electrons and molecular orbitals 
in the individual irreducible representations of the $\mathrm{C_{2v}}$ 
point group were $\mathrm{A_1}\rightarrow2/5$, $\mathrm{B_{1}}\rightarrow0/2$, 
$\mathrm{B_{2}}\rightarrow0/2$, $\mathrm{A_{2}}\rightarrow0/0$. Using 
these values we could achieve reasonable agreement with other results 
in the literature \cite{JCP1,JCP2}.

\subsection{The working Hamiltonian}

\begin{figure}[!ht]
 \centering
 \includegraphics[width=0.4\textwidth]{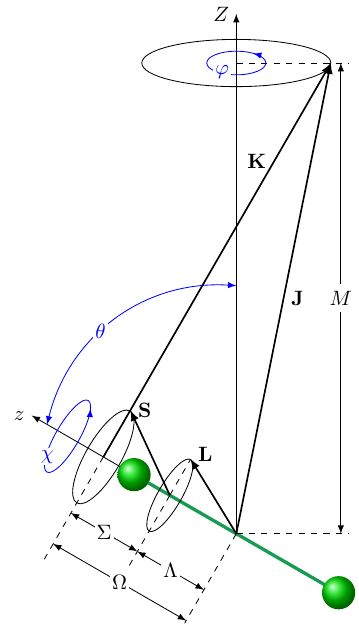}
\caption{Vector diagram of the electron-rotation coupling conforming to Hund's case (a). 
$\Lambda$ and $\Sigma$ are the projections of the electronic orbital $\mathbf{L}$ 
and spin $\mathbf{S}$ angular momenta onto the molecular axis, i.e. the $z$ axis 
of the molecule-fixed (MF) frame. $\mathbf{K}$ is the nuclear rotational angular momentum, 
and $\mathbf{J}$ is the total angular momentum with $\mathrm{M}$ being its projection onto 
the space fixed (SF) $Z$ axis. $\varphi$ and $\chi$ are the Euler angles representing 
rotations around the SF $Z$ and MF $z$ axes, while $\theta$ is the angle between these 
two.}\label{fig:fig2}
\end{figure}

As mentioned above, in the present work we extend our previously used 
model for light-matter interaction to include the $\mathbf{K}$-$\Omega$ 
coupling. A schematic representation of this scenario in Hund's case (a) 
is presented in Fig. \ref{fig:fig2}. In this scheme, both the electronic 
orbital angular momentum $\mathbf{L}$ and spin angular momentum $\mathbf{S}$ 
are tied to the molecular axis with projections $\Lambda$ and $\Sigma$ 
on it. The total angular momentum $\mathbf{J}$ of the system is the vector 
sum of this total interior angular momentum and the rotational angular 
momentum $\mathbf{K}$. Accordingly, the projection of $\mathbf{J}$ onto 
the molecular axis, i.e. the body-fixed (BF) $z$ axis, and the space-fixed 
(SF) $Z$ axis are $\Omega=\Lambda+\Sigma$ and $\mathrm{M}$, respectively.

Molecular rotations are usually described in theoretical investigations in the basis of 
spherical harmonics or Legendre polynomials. However, this convention is unable to treat 
the $\mathbf{K}$-$\Omega$ coupling. For this reason, the more general Wigner D basis 
\begin{align}
 |JM\Omega\rangle = \sqrt{\frac{2J+1}{8\pi^2}} D_{M\Omega}^{J}(\varphi,\theta,\chi) 
 = \sqrt{\frac{2J+1}{8\pi^2}} \, \mathrm{e}^{-iM\varphi}\, d_{M\Omega}^{J}(\theta)\, \mathrm{e}^{-i\Omega\chi}
\end{align}
is needed, where the Wigner D-functions $D_{M\Omega}^{J}(\varphi,\theta,\chi)$ are the 
matrix elements of the rotation operator $\mathbf{D}(\varphi,\theta,\chi)$. They may be 
represented as a product of three functions, each depending on a single argument, with 
$d_{M\Omega}^{J}(\theta)$ a real function known as the reduced Wigner D-function. The 
arguments $\varphi$ and $\chi$ represent Euler angles of rotation around the SF $Z$ 
and BF $z$ axes, while $\theta$ is the Euler angle between these two axes, as presented 
in Fig. \ref{fig:fig2}.

The time-dependent nuclear Hamiltonian of the $\mathrm{NaH}$ molecule in the 
space of the three lowest singlet electronic states interacting with a linearly 
polarized laser field $\mathcal{E}_{Z}(t)$ may be written as
\begin{align}
 \mathbf{H}(\mathbf{R}, t) =& \left(-\frac{1}{2M_{r}}\frac{\partial^{2}}{\partial R^2} + \frac{\mathbf{K}^2}{2M_{r}R^2} \right)\cdot \mathbf{1} +
 \begin{pmatrix}
  V_{X}(R) & 0 & 0\\
  0 & V_{A}(R) & 0\\
  0 & 0 & V_{B}(R)
 \end{pmatrix}-\nonumber\\
 &\hspace*{1cm}-\mathcal{E}_{Z}(t)
 \begin{pmatrix}
  \mu_{XX}\,\kappa_{XX} & \mu_{AX}\,\kappa_{AX} & \mu_{BX}\,\kappa_{BX}\\
  \mu_{AX}\,\kappa_{AX}^{\dagger} & \mu_{AA}\,\kappa_{AA} & \mu_{BA}\,\kappa_{BA}\\
  \mu_{BX}\,\kappa_{BX}^{\dagger} & \mu_{BA}\,\kappa_{BA}^{\dagger} & \mu_{BB}\,\kappa_{BB}
 \end{pmatrix},\label{Eq:Hamilton_working}
\end{align}
where the first term describes the vibrational and rotational kinetic energy 
operator for the reduced mass $M_{r}$ and internuclear distance $R$, and 
$\mathbf{1}$ is the three dimensional unit matrix. The second term stands for 
the potential energy operator with the potential energy curves $V_{X/A/B}(R)$ 
of the considered electronic states. The interaction term contains both the 
PDMs in the diagonal, and the TDMs in the off-diagonal elements, with 
\begin{align}
 \kappa_{f\,i}(\theta,\chi)=
 \begin{dcases}
  \cos\theta & \text{for }\Delta\Lambda_{f \leftarrow i} = 0\\
  -\sin\theta\mathrm{e}^{\pm i\chi} & \text{for }\Delta\Lambda_{f \leftarrow i} = \pm 1.
 \end{dcases}\label{Eq:kappa_if}
\end{align}
Details of the derivation of this interaction term can be found in the Appendix \ref{Sec:Appendix1}.

The working Hamiltonians used in our 1D and 2D calculations are obtained 
as a particular form of the above expression by setting $\Omega=0$, and 
correspondingly $\chi=0$. In this case the rotational angular momentum 
coincides with the total angular momentum of the system, 
$\mathbf{K}\equiv\mathbf{J}$, and the normalized Wigner D basis reduces 
the spherical harmonics, $|JM\Omega\rangle \to |JM\rangle \equiv Y_{J}^{M}(\theta,\varphi)$. 
Moreover, the $M$-selection rule remains valid, so we recover our usual 
formalism \cite{Attila2, Attila3} in terms of the Legendre polynomials 
$P_{J}(\theta)$. Additionally, in the 1D model the rotational kinetic 
energy is completely neglected ($\mathbf{K}$=0) \cite{Gabor3, Gabor4}.

\subsection{The applied electric field }

In the present work we investigated the photodissociation process of 
the $\mathrm{NaH}$ molecule in a pump-probe setup. Accordingly, the 
electric field $\bm{\mathcal{E}}$ entering the interaction term of 
our working Hamiltonian (\ref{Eq:Hamilton_working}) is the sum of 
two linearly polarized pulses in the $Z$-direction
\begin{align}
 \bm{\mathcal{E}}(t) \equiv \bm{\mathcal{E}}_{Z}^{SF}(t) = 
 \mathbf{e}_{Z}\,\mathcal{E}_{pm}\,f_{pm}(t)\cos\left(\omega_{pm} t \right) + 
 \mathbf{e}_{Z}\,\mathcal{E}_{pr}\,f_{pr}(t)\cos\left(\omega_{pr} t \right),
\end{align}
with amplitudes $\mathcal{E}_{i}$ and energy $\hbar\omega_{i}$.
Both pulses were considered with cos-square-shaped envelope functions
\begin{align}
 f_{i}(t) = \cos^2\left( \frac{\pi(t-t_{i})}{\tau_{i}} \right) \qquad \text{for $i\in\{pm, pr\}$},
\end{align}
where $\tau_{pm}$ and $\tau_{pr}$ are the durations of the pump and 
the probe pulse, respectively. The pump pulse, centered at $t_{pm}=0$ fs 
(hence defining our time axis), initiates the dynamics of the system 
by transferring some part of the initial ground $X$ state population 
to the first excited $A$ state. The time delayed probe pulse, 
$t_{pr}=t_{pm}+\Delta t$ with $\Delta t=-25\ldots1000$ fs, can either 
transfer population back from $A$ to the $X$ state or further promote 
to the dissociative $B$ state.

\begin{figure}[!ht]
 \centering
 \includegraphics[width=0.99\textwidth]{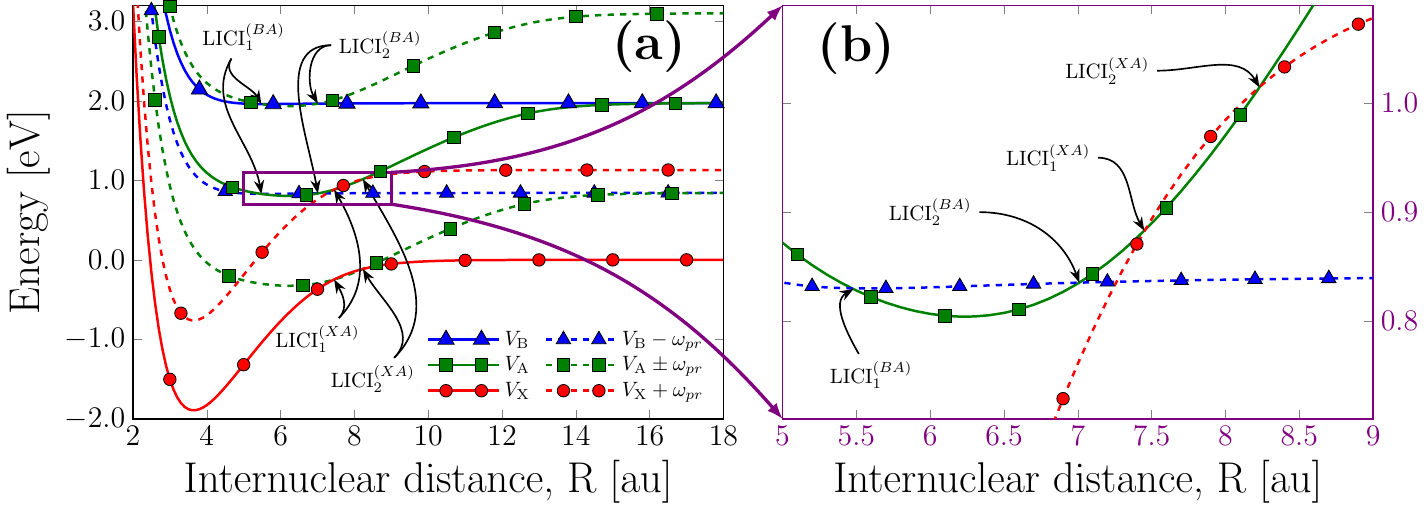}
\caption{(a) The three lowest-lying singlet adiabatic potential
energy curves of the $\mathrm{NaH}$ molecule. The corresponding light-dressed
states are denoted by dashed lines. The position of the light-induced
conical intersections ($\mathrm{LICI}_{1}^{(BA)}$, $\mathrm{LICI}_{2}^{(BA)}$, 
$\mathrm{LICI}_{1}^{(XA)}$ and $\mathrm{LICI}_{2}^{(XA)}$) are also marked. 
(b) Magnification of the region highlighted on panel (a) with the light-induced 
conical intersections.
}\label{Fig:fig3}
\end{figure}

As mentioned above, the position of light-induced conical intersections 
can be controlled by tuning the laser energy. In our previous study on 
the $\mathrm{NaH}$ molecule \cite{Umarov1}, the probe pulse was able to 
produce only a light induced avoided crossing between the $X$ and $A$ 
states. Here, we slightly increased its energy to $\hbar\omega_{pr}=1.13$ eV 
which lead to two conical intersections ($\mathrm{LICI}_{1}^{(XA)}$ 
and $\mathrm{LICI}_{2}^{(XA)}$) between these surfaces, as can be seen 
on panel (a) of Fig. \ref{Fig:fig3}. The pump energy was set as the 
third-harmonic of this frequency, $\hbar\omega_{pm}=3.39$ eV. For the 
pulse duration of the pump we chose $\tau_{pm}=22$ fs, keeping its 
spectral width narrow enough to minimise population transfer to the 
$B$ state. The probe pulse was $\tau_{pr}=30$ fs long. With a pump 
intensity of $I_{pm}=1\times10^{12}$ $\mathrm{W/cm^2}$ we obtained 
meaningful excitation ($\sim 35\%$ on the $A$ state) without Rabi 
floppings between $X$ and $A$. In our numerical simulations we 
investigated two intensity regimes of the probe pulse. In the low 
intensity case of $I_{pr}=1\times 10^{11}$ $\mathrm{W/cm^2}$ one-photon 
processes dominate, while for the higher intensity of 
$I_{pr}=1\times 10^{12}$ $\mathrm{W/cm^2}$ multiphoton processes may occur, 
and fingerprints of the electron-rotation coupling are expected to manifest. 

In order to illustrate the heart of the LICI concept, we employ the 
well-known Floquet framework (see \cite{Gabor1,Gabor2,Gabor3,Gabor4} 
for a detailed discussion), which is commonly used to understand various 
phenomena in light-matter physics. The action of the probe pulse is 
described in this picture as the upward and downward shift of the 
potential energy curves by integer multiples of photon energy $n\hbar\omega_{pr}$. 
This is illustrated in Fig. \ref{Fig:fig3} (panels (a) and (b)) for 
the most basic $n=1$ one-photon case. As a result, the upward shifted 
first excited $V_{A}(R)+\hbar\omega_{pr}$ curve intersects the second 
excited $V_{B}(R)$ curve (or equivalently $V_{A}(R)$ is intersected 
by the downward shifted $V_{B}(R)-\hbar\omega_{pr}$), creating two 
light-induced conical intersections $\mathrm{LICI}_{1}^{(BA)}$ and 
$\mathrm{LICI}_{2}^{(BA)}$. The same phenomena appears for 
$V_{A}(R)-\hbar\omega_{pr}$ and $V_{X}(R)$ (identically between 
$V_{X}(R)+\hbar\omega_{pr}$ and $V_{A}(R)$) leading to 
$\mathrm{LICI}_{1}^{(XA)}$ and $\mathrm{LICI}_{2}^{(XA)}$. 
Diagonalizing the corresponding Floquet form of the Hamiltonian 
Eq. (\ref{Eq:Hamilton_working}), one can obtain the light-induced 
adiabatic potential energy surfaces. The resulting conical intersections 
$\mathrm{LICI}^{(fi)}$ between two states $i,f\in\{X, A, B\}$ appear in 
the configuration space whenever the conditions $V_{i}(R)\pm\hbar\omega_{pr}=V_{f}(R)$ 
(specifying its $R$-coordinate) and $\kappa_{fi}(\theta)=0$ (see Eq. (\ref{Eq:kappa_if})) 
are simultaneously fulfilled. For parallel (i.e. $\Sigma \leftrightarrow \Sigma$ 
or $\Pi \leftrightarrow \Pi$) transitions this occurs at $\theta=\pi/2$, 
while for perpendicular ones ($\Sigma \leftrightarrow \Pi$) at $\theta=0$ 
and $\theta=\pi$.

\subsection{Nuclear wave packet propagation and calculated quantities}

The nonadiabatic nuclear dynamics of the $\mathrm{NaH}$ molecule in the 
LICI framework was investigated by solving the time-dependent Schr\"{o}dinger 
equation (TDSE) governed by the Hamiltonian of Eq. (\ref{Eq:Hamilton_working}). 
Our simulations were performed with the multiconfigurational time-dependent 
Hartree (MCTDH) method implemented in the Heidelberg package \cite{MCTDH1,MCTDH2,MCTDH3}. 
The vibrational degree of freedom ($R$) was discretized by a Sin-DVR 
(sine discrete variable representation) with $N_R=512$ grid points distributed 
in the internuclear coordinate range $[2.0, 50.0]$ au. The rotational 
degree of freedom ($\theta$) was expanded in the basis of Legendre 
polynomials for the conventional description, and in the basis of 
$\mathrm{L}_2$ normalized Wigner D-functions when the $\mathbf{K}$-$\Omega$ 
electron-rotation coupling was included. In both cases $N_{\theta}=91$ 
basis functions were employed. The Wigner D basis was further constructed 
with $N_{\chi}=11$ and $N_{\varphi}=1$ exp-DVR functions for $\chi$ and 
$\varphi$, respectively. In the MCTDH formalism, these so called primitive 
basis sets ($\eta$) are used to represent the single particle functions (SPF)
\begin{align}
 \phi_{j_{q}}^{(q)}(q,t) = 
 \begin{dcases}
  \sum_{l=1}^{N_{q}}c_{j_{q}\,l}^{(q)}(t) \,\eta_{l}^{(q)}(q), & q \in\{R, \theta\},\\
  \sum_{l_{\theta}=1}^{N_{\theta}} \sum_{l_{\varphi}=1}^{N_{\varphi}} \sum_{l_{\chi}=1}^{N_{\chi}} 
  c_{j_{q}\,l_{\theta}\,l_{\varphi}\,l_{\chi}}^{(q)}(t) \prod_{f\in\{\theta,\varphi,\chi\}} \eta_{l_{f}}^{(f)}(f), & q=(\theta,\varphi,\chi).
 \end{dcases}
\end{align}
The second expression applies in the $\mathbf{K}$-$\Omega$ case, when 
the primitive basis functions of the $\theta$, $\varphi$ and $\chi$ 
degrees of freedom are combined into a single MCTDH mode to form 
three-dimensional single-particle functions. The SPFs are used in 
turn to build up the wave function
\begin{align}
  \psi(R, \rho, t) &= \sum_{j_{R}}^{n_{R}}\sum_{j_{\rho}}^{n_{\rho}} A_{j_{R},j_{\rho}}(t) \,
  \phi_{j_{R}}^{(R)}(R,t)\, \phi_{j_{\rho}}^{(\rho)}(\rho,t), \qquad \text{where }\rho\in\{\theta,\, (\theta,\varphi,\chi)\}.
\end{align}
Throughout our simulations, the single-particle basis size was
$n_{R}=n_{\rho}=20$ for both modes in the expansion of the wave 
function, which ensured the convergence of the results.

A complex absorbing potential (CAP) of the form 
$-iW(R)=-i\eta(R-R_{c})^{3}\Theta(R-R_{c})$ is applied to the 
vibrational degree of freedom on all three surfaces in order 
to keep the length of its primitive grid manageable while 
preventing spurious reflections. The starting point and 
strength of the CAP were $R_c=40$ [au] and $\eta=1.2\times10^{-5}$, 
respectively, while the Heaviside's $\Theta$ function means 
that it affects the $R>R_c$ region. It was also used to perform 
flux analysis \cite{MCTDH3} to obtain physical quantities of 
the photodissociation process such as the kinetic energy release 
spectra (KER) or the photofragment angular distribution (PAD) 
\begin{align}
 P_{\mathrm{KER}}(E) &= \int_{0}^{\infty}\mathrm{d}t \int_{0}^{\infty}\mathrm{d}t' \, \left\langle \psi(t) \left| W \right| \psi(t') \right\rangle \mathrm{e}^{-iE(t-t')},\label{Eq:KER}\\
 P_{\mathrm{PAD}}(\theta_{j}) &= \frac{1}{w_j}\int_{0}^{\infty}\mathrm{d}t \, \left\langle \psi(t) \left| W_{\theta_{j}} \right| \psi(t) \right\rangle. \label{Eq:PAD}
\end{align}
In the above expression of the PAD, $\theta_{j}$ and $w_{j}$ 
are the specific angles and their corresponding quadrature 
weights in the applied DVR, while $-iW_{\theta_j}$ is the 
projection of the CAP onto these grid points. Strictly speaking, 
the angular distribution of Eq. (\ref{Eq:PAD}) gives the 
probability of photodissociation in the solid angle 
$(\Omega,\Omega+\mathrm{d}\Omega)$ in the direction $(\theta, \varphi)$. 
Due to the cylindrical symmetry of the problem, also exploited 
when choosing a Legendre basis instead of spherical harmonics 
for the rotational degree of freedom in our 1D and 2D models, 
it is independent of $\varphi$. Alternatively, we may calculate 
the total photofragment angular distribution in the $\theta_j$ 
direction by integrating Eq. (\ref{Eq:PAD}) along $\varphi$ as
\begin{align}
 \tilde{P}_{PAD}(\theta_j) = \int_{0}^{2\pi}P_{PAD}(\theta_j, (\varphi))\sin\theta_j\,\mathrm{d}\varphi
 \propto \int_{0}^{\infty}\mathrm{d}t \, \left\langle \psi(t) \left| W_{\theta_{j}} \right| \psi(t) \right\rangle, \label{Eq:PAD_tot}
\end{align}
where in the final experssion we used the fact that the quadrature 
weights are proportional to $\sin\theta_j$.

The total dissociation probability is easily obtained by integrating 
either of the above quantities 
\begin{align}
 P_{\mathrm{diss}} = \int_{0}^{\infty} P_{\mathrm{KER}}(E)\, \mathrm{d}E 
 = \int_{0}^{\pi}P_{\mathrm{PAD}}(\theta) \sin\theta\, \mathrm{d}\theta 
 = 2\int_{0}^{\infty}\mathrm{d}t\left\langle \psi(t) \left| W \right| \psi(t) \right\rangle.\label{Eq:Pdiss}
\end{align}

To demonstrate the impact of light-induced nonadiabatic effects 
on the dissociation dynamics of $\mathrm{NaH}$ we compare results 
obtained with two levels of theory relative to the rotational 
degree of freedom $\theta$. In the full two-dimensional (2D) case 
$\theta$ is considered as a dynamical variable, while in the 
restricted one-dimensional (1D) situation it is only a parameter. 
This means that the original orientation of the molecules remain 
unchanged during the evolution of the system, and they experience 
an effective field strength $\mathcal{E}^{eff}(\theta)=\mathcal{E}_{pm,pr}\kappa_{fi}(\theta)$, 
corresponding to the effective intensity $I_{0}^{eff}(\theta)=I_{0}\cos^2\theta$ 
or $I_{0}^{eff}(\theta)=I_{0}\sin^2\theta$, through the permanent 
and transition dipoles depending on the symmetry of the involved 
electronic states. The difference of the two descriptions is that 
the degeneracy points between the potential energy surfaces manifest 
as LICIs in 2D and LIACs in 1D. Our third type of simulations, 
referred as 3D, explore the impact of electron-rotation coupling 
in the LICI picture.

\section{Results and discussion }

The aim of this work is to describe and characterize the photodissociation 
dynamics of a nonaligned, isotropically distributed NaH molecule in the 
presence of light-induced conical intersections, and to explore the effect 
of the usually neglected electron-rotation coupling on this process. To 
this end, we compute dissociation probabilities, kinetic energy release (KER) 
spectra, and photofragment angular distributions (PAD) in a pump-probe setup.

\begin{figure}[!ht]
 \centering
 \includegraphics[width=0.95\textwidth]{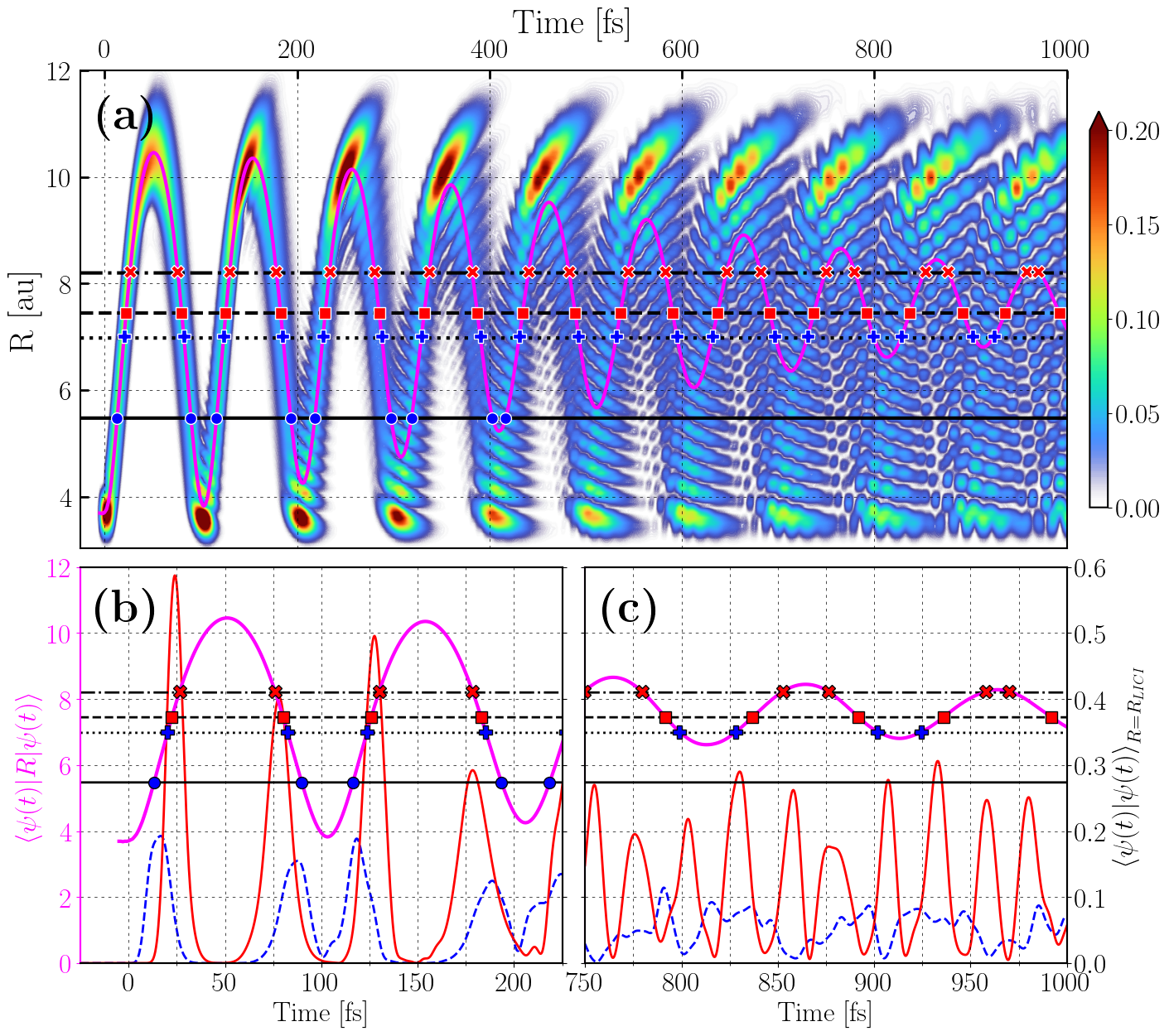}
 \caption{(a) Vibrational motion of the nuclear wave packet on the A state 
 after excitation by the pump pulse, and the expectation value of  the $R$ 
 coordinate (magenta curve). The horizontal lines show the LICI positions, 
 while the markers indicate when the $R$-expectation curve crosses them: 
 $\mathrm{LICI}_{1}^{(BA)}$ continuous line / blue circles; 
 $\mathrm{LICI}_{2}^{(BA)}$ dotted line / blue plus signs; 
 $\mathrm{LICI}_{1}^{(XA)}$ dashed line / red squares; and $\mathrm{LICI}_{2}^{(XA)}$ 
 dash-dotted line / red $\times$ signs. Panels (b) and (c) zoom in the two 
 investigated delay time windows, and show the $R$-expectation curve and 
 LICI position indicators as on panel (a), along the TDM weighted sums of 
 the nuclear wavepacket density at $\mathrm{LICI}_{1+2}^{(BA)}$ with dashed 
 blue line, and $\mathrm{LICI}_{1+2}^{(XA)}$ with continuous red line.}\label{Fig:den1d}
\end{figure}

The system is assumed to be initially prepared in its electronic 
($\mathrm{X}$), vibrational ($\nu=0$), and rotational ($J=0$) ground state. 
The dynamics is initiated by the pump-pulse, which promotes about 35\% of 
the population to the $\mathrm{A}$ state. The excited wave packet starts 
to oscillate within an interval of $R\approx3-12$ a.u.. This field-free 
(without a probe pulse) time evolution is displayed in Fig. \ref{Fig:den1d} a). 
A reasonably good indicator for the instantaneous position of the wave packet 
is the expectation value of the internuclear distance 
$\left\langle \psi(t) \left| R \right| \psi(t) \right\rangle$. 
This quantity is overlayed on top of the vibrational density map 
(integrated over the angular degrees of freedom) of Fig. \ref{Fig:den1d} a) 
with a continuous magenta curve. During the initial stages, within the 
first period of the vibrational motion, the wavepacket is well localized, 
and the $R$-expectation value follows it almost exactly. As time progresses, 
due to the multiple vibrational components with different energies (and 
periods), the wavepacket becomes increasingly more diffuse, therefore the 
$R$-expectation curve is only able to reflect the overall periodicity of 
the motion.

The horizontal lines in Fig. \ref{Fig:den1d} indicate the internuclear 
distances where light induced conical intersections are formed by the 
probe pulse, as in Fig. \ref{Fig:fig3}. Specifically, the continuous 
and the dotted lines at $R=5.48$ [a.u.] and $R=7.01$ [a.u.] indicate 
the intersections of the $V_{B}(R)$ and $V_{A}(R)$ potential energy 
curves, that is $\mathrm{LICI}_{1,2}^{(BA)}$. Similarly, the dashed 
and dash-dotted lines mark the $\mathrm{LICI}_{1,2}^{(XA)}$ positions 
at $R=7.45$ [a.u.] and $R=8.23$ [a.u.] between the $V_{X}(R)$ and $V_{A}(R)$ 
curves. Dissociation in the individual electronic channels is expected 
to take place around those probe delay times when the vibrational 
wavepacket traverses the interaction regions. These are marked for 
the dissociative $\mathrm{B}$ state with blue circles and blue plus 
signs for $\mathrm{LICI}_{1}^{(BA)}$ and $\mathrm{LICI}_{2}^{(BA)}$, 
respectively. For later times the $R$-expectation curve does not reach 
the $\mathrm{LICI}_{1}^{(BA)}$ distance, but the wavepacket clearly 
does. For population with sufficiently high kinetic energy dissociation 
can occur on state $\mathrm{X}$ at $\mathrm{LICI}_{1}^{(XA)}$ and 
$\mathrm{LICI}_{2}^{(XA)}$ marked with red squares and red $\bm{\times}$ 
signs, respectively.

Panels b) and c) of Fig. \ref{Fig:den1d} highlight the two time windows 
of the above dymanics we focus on in the rest of this work. Here, the 
$R$-expectation value and the LICI indicators (horizontal lines and 
markers) are reproduced on the left $y$-axis. The data belonging to 
the right $y$-axis shows the time evolution of the nuclear wavepacket 
density at the crossings to state $\mathrm{B}$ with dashed blue line 
and $\mathrm{X}$ with continuous red line. To be more exact, the 
vibrational densities at the internuclear distances ($R_{LICI_{1}^{(fA)}}$ 
and $R_{LICI_{2}^{(fA)}}$) corresponding to crossings between the same 
two potential energy curves are weighted with the appropriate transition 
dipole values and summed together. This quantity may serve as a crude 
indicator of the dissociation yield in the corresponding channels.

\subsection{Total Dissociation probability}

\begin{figure}[!ht]
 \centering
 \includegraphics[width=0.7\textwidth]{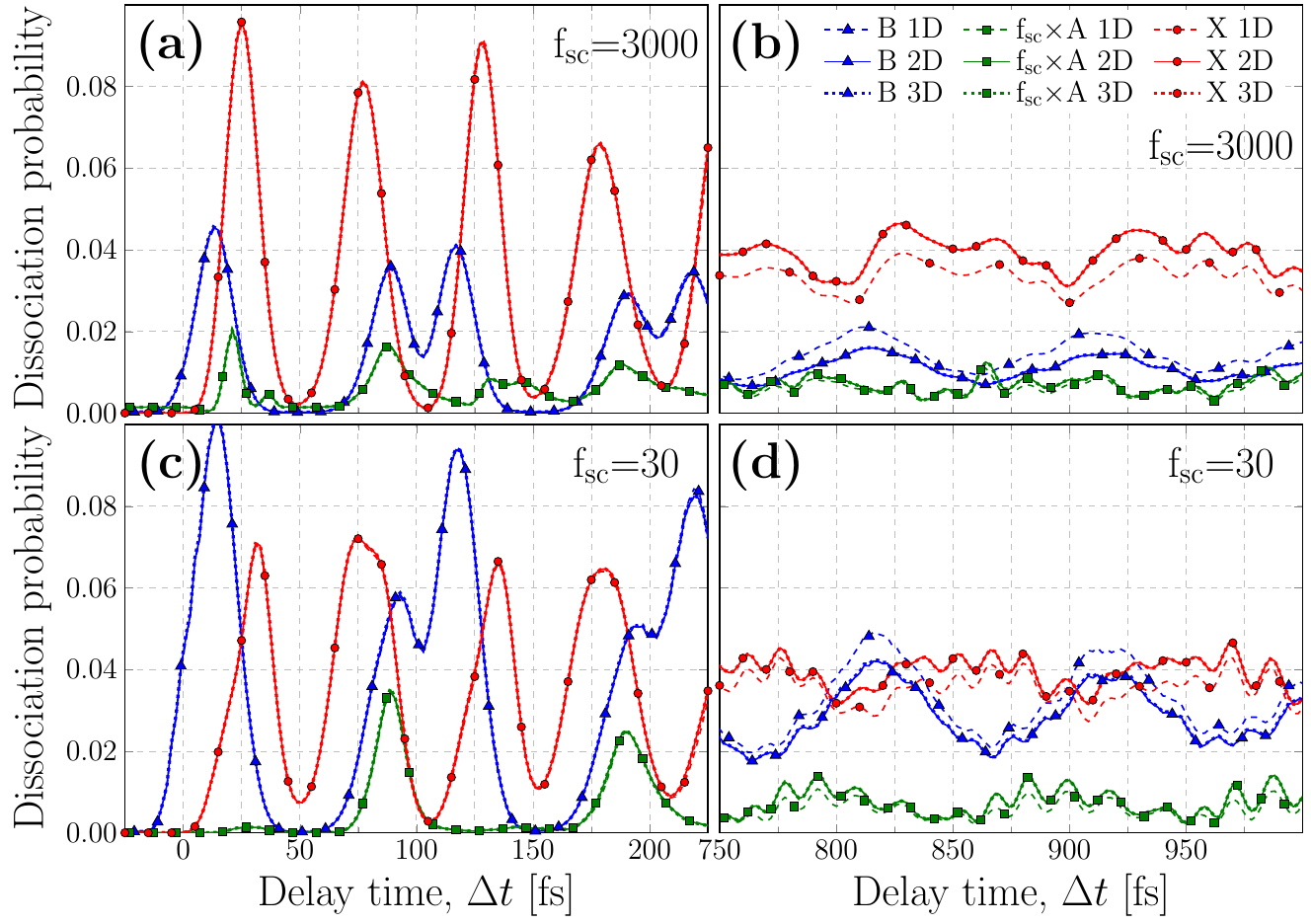}
 \caption{Dissociation probabilities corresponding to the individual 
 adiabatic electronic states  ($\mathrm{X}$ - red with circles, 
 $\mathrm{A}$ - green with squares, $\mathrm{B}$ - blue with triangles)
 as a function of delay time between the pump and the probe pulses. 
 The $\mathrm{A}$ state dissociation  probability is scaled for 
 better visibility, with the scaling factor $\mathrm{f_{sc}}$ 
 indicated on each panel. The dashed, continuous and dotted lines 
 correspond to results obtained with the 1D, 2D and 3D models, respectively.
 Panels (a) and (b) show our results for $I_{pr}=1\times10^{11}$ $\mathrm{W/cm^2}$, 
 while (c) and (d) for $I_{pr}=1\times10^{12}$ $\mathrm{W/cm^2}$}\label{Fig:Pdiss}
\end{figure}

The total dissociation yields on the individual electronic states 
($\mathrm{X}$: red, circle; $\mathrm{A}$: green, square; 
$\mathrm{B}$: blue, triangle) calculated with Eq. (\ref{Eq:Pdiss}) 
within the 1D (dashed), 2D (continuous) and 3D (dotted) models are 
presented in Fig. \ref{Fig:Pdiss}. The upper row (panels a) and b)) 
show the results obtained in the two delay-time windows for 
$I_{pr}=1\times 10^{11}$ $\mathrm{W/cm^2}$, while the lower row (c) 
and d)) for the higher employed pump intensity 
$I_{pr}=1\times 10^{12}$ $\mathrm{W/cm^2}$. In case of the low 
$I_{pr}$ the probability curves corresponding to the $\mathrm{B}$ 
and $\mathrm{X}$ states mimic nicely the density curves presented 
in Fig. \ref{Fig:den1d} b) and c). This indicates that one-photon 
transition is the dominant mechanism, and in this range the 
dissociation yields scale linearly with the intensity. In the small 
$\Delta t$ window the dissociation probabilities show well separated 
peaks which align perfectly with the transition times defined by the 
intersection of the $R$-expectation curve and the LICI lines. For 
large delay times we do not get such well defined structures due to 
the blurring of the excited wavepacket. For the higher intensity the 
result for state $\mathrm{B}$ roughly double in value, which shows 
that we are still in the linear regime, but the peaks are distorted 
due to saturation. In contrast, dissociation probability in the 
$\mathrm{A}$ channel does not increase, instead certain peaks even 
get diminished. This is explained by the fact that the vibrtional 
wavepacket encounters the $\mathrm{B}$ crossings first, and as a 
result by the time it reaches $\mathrm{LICI}_{1,2}^{(XA)}$ there is 
less population available for transition. For large $\Delta t$ 
values the dissociation yield curves become more structured, which 
reflects the interference pattern of the different wave packet 
vibrational components seen in the density map of Fig. \ref{Fig:den1d} a). 
Figure \ref{Fig:Pdiss} contains a third group of data, shown with 
green lines with squares. These represent dissociation on the 
$\mathrm{A}$ state. This is inherently a two-photon process, population 
transfer from $\mathrm{A}$ to $\mathrm{B}$ and/or $\mathrm{X}$ followed 
by transfer back to $\mathrm{A}$, which has a lower probability. For 
this reason it has to be scaled up to be visible on the same scale as 
the other two dissociation probabilities. The ratio of the scaling 
factors, $\mathrm{f_{sc}}$=3000 for the low and $\mathrm{f_{sc}}$=30 
for the high intensity, and the approximate doubling of its value for 
an order of magnitude increase of the intensity confirms this mechanism.

Comparing the results obtained with the three approaches, we see 
that for small delay times they essentially coincide. For large 
$\Delta t$ the 2D and 3D data remain indistinguishable, but the 
1D curves deviate noticeably. In the $\mathrm{X}$ channel it 
consistently lie below the 2D/3D curves, while in the $\mathrm{B}$ 
channel it yields higher dissociation values for all delay times. 
This is true for both employed probe intensities. These observations 
are jointly determined by molecular rotation and the emerging 
light-induced nonadiabatic effects. The pump pulse initiates the 
rotational motion of the molecule. However, this is a relatively 
slow process and its influence is not yet apparent at short delay 
times. As a consequence, in this domain, the dissociation probabilities 
calculated with the three models are identical. For later times, 
the rotation manifests in the alignment of the molecules along the 
laser polarization, i.e. $Z$-axis. This means, when the probe pulse 
arrives, it interacts with more favorably oriented molecules than 
in the 1D approach, where rotation is neglected. Due to the symmetries 
of the electronic states, this alignment enhances the coupling between 
the $\mathrm{A}\leftrightarrow\mathrm{X}$ states and reduces the 
one between $\mathrm{A}\leftrightarrow\mathrm{B}$. This explains the 
trends observed between the results obtained with the models that 
include rotational motion (2D, 3D) and the 1D that neglects it. For 
this physical quantity, the incorporation of electron-rotation in 
the description does not show any effect.

\subsection{Kineti Energy Release Spectra}

\begin{figure}[!ht]
 \centering
 \includegraphics[width=0.9\textwidth]{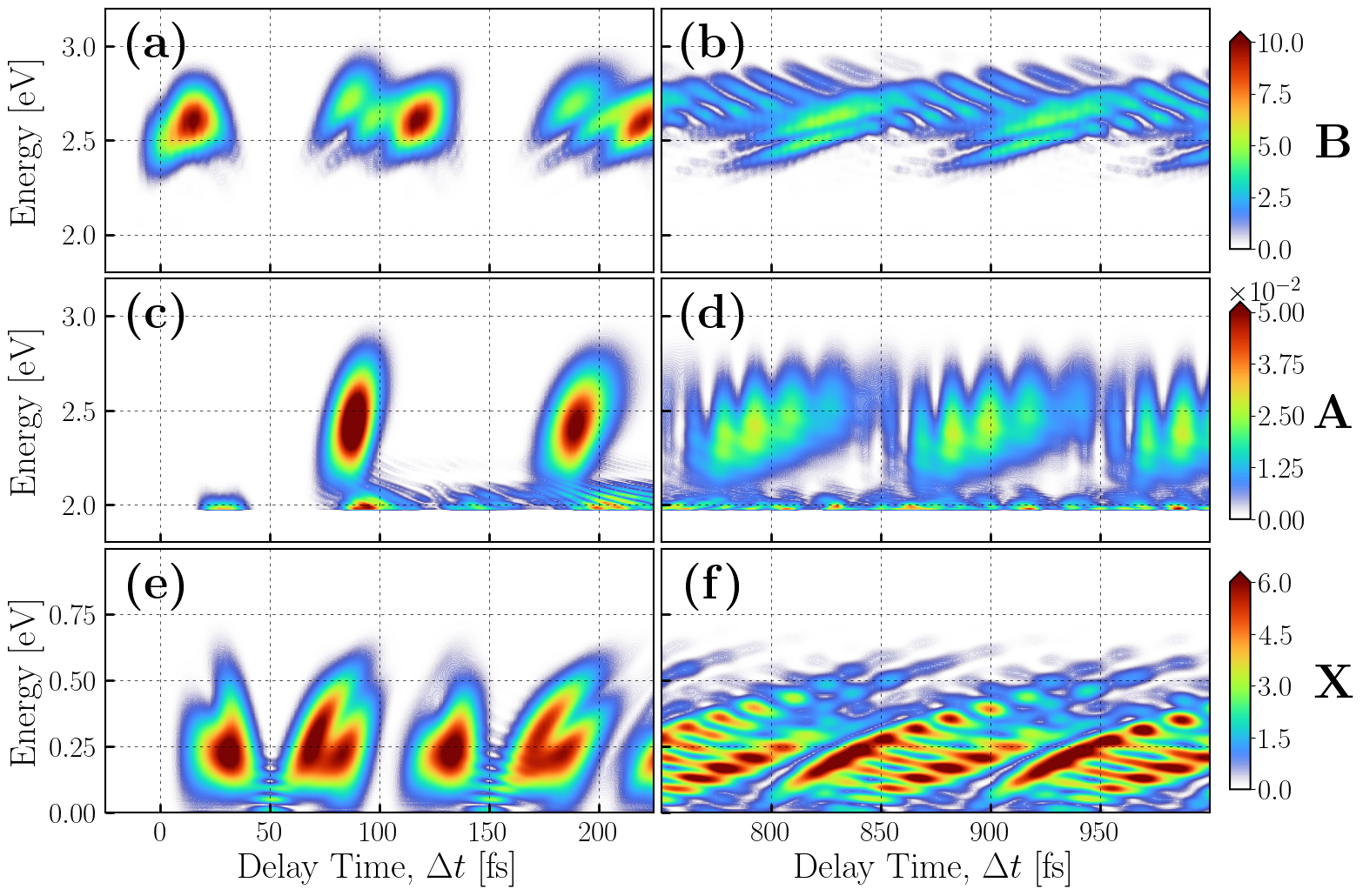}
 \caption{Kinetic energy release (KER) spectra of the photofragments 
 for $I_{pr}=1\times 10^{12}$ $\mathrm{W/cm^2}$ as a function of 
 delay time for the 2D model on each dissociation channel: $\mathrm{B}$ 
 state (a, b); $\mathrm{A}$ state (c, d); $\mathrm{X}$ state (e, f).}\label{Fig:KER}
\end{figure}

To illustrate the time-delay dependence of the kinetic energy release 
spectra of the photofragments, we present in Fig. \ref{Fig:KER} the 
results obtained with the 2D model at the higher investigated intensity. 
Panels in different rows correspond to spectra of the individual 
electronic states. At short delay times, shown in left column, the 
spectra exhibit well defined peaks around the time moments the excited 
wave packet on the $\mathrm{A}$ state crosses one of the LICIs. 
In contrast, for long delay times as the different vibrational 
components of the excited wave packet do not reach the coupling regions 
simultaneously, more elongated and structured features emerge.

From our previous work \cite{Umarov1} it is already known, that the 
1D and 2D models yield nearly identical KER spectra at short delay 
times. At longer delays, however, significant differences emerge due 
to the combined effect of the rotational motion initiated by the pump 
pulse and the light-induced nonadiabatic dynamics. These mechanisms 
have been discussed above describing the differences observed in the 
dissociation yields calculated with the different methods. 

\begin{figure}[!ht]
 \centering
 \includegraphics[width=0.85\textwidth]{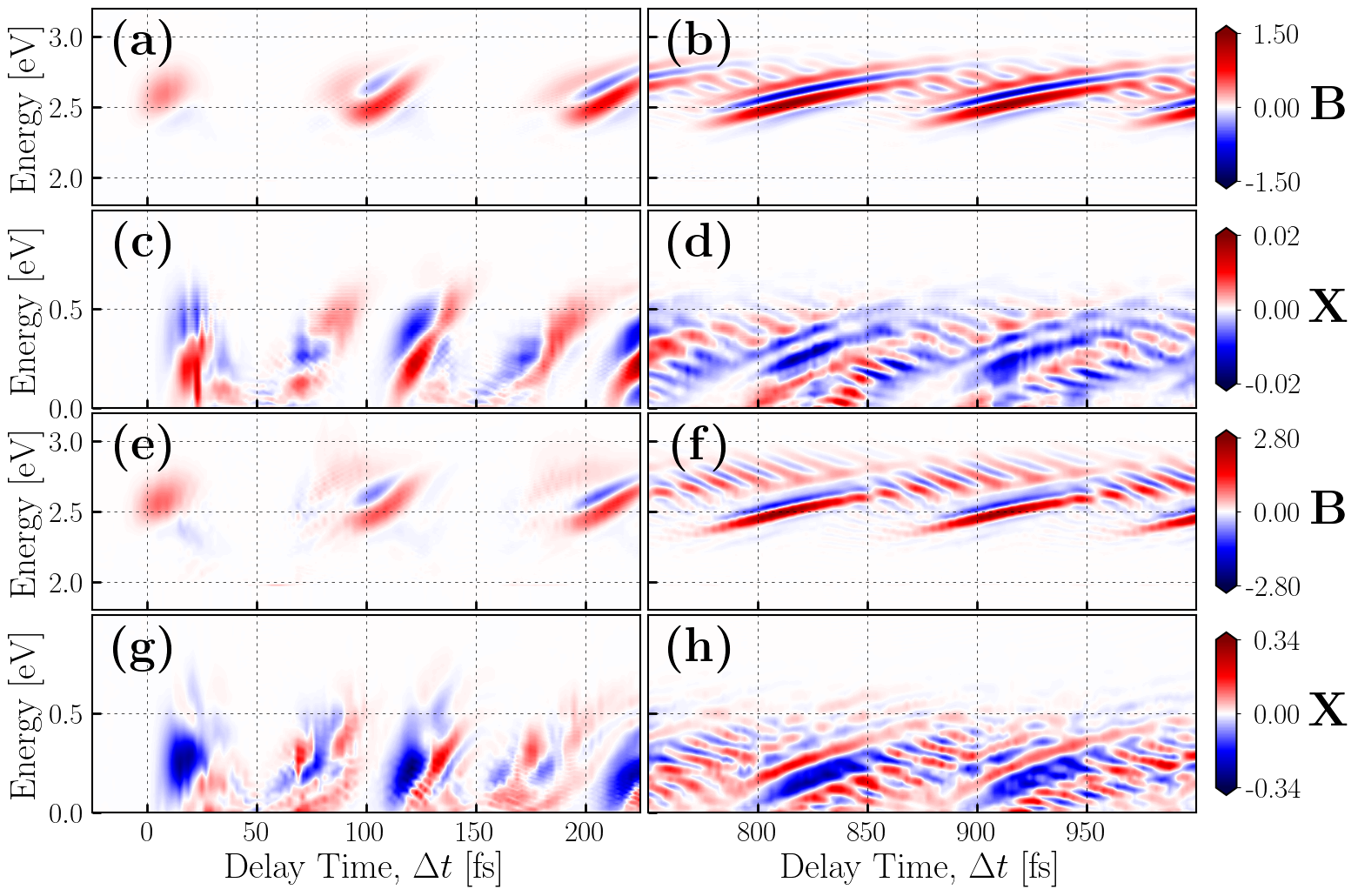}
 \caption{Relative difference between the KER results obtained 
 in the 2D and 3D models for the B and X dissociation channels. 
 The color scale represents percentages.  Panels (a)-(d) show 
 the results for $I_{pr}=1\times 10^{11}$ $\mathrm{W/cm^2}$, 
 while (e)-(h) for $I_{pr}=1\times 10^{12}$ $\mathrm{W/cm^2}$, 
 respectively.}\label{Fig:KER_diff}
\end{figure}
The primary focus of the present study is to highlight possible 
effects of the electron-rotation coupling that the simpler models 
might neglect. For this reason, we present in Fig. \ref{Fig:KER_diff} 
the relative difference between the 2D and 3D KER spectra on the 
$\mathrm{X}$ and $\mathrm{B}$ states for both investigated probe-pulse 
intensities. The spectra on the $\mathrm{A}$ state is orders of 
magnitude smaller than on the other two channels, as indicated by 
the scale of the colorbars of Fig. \ref{Fig:KER}, therefore they 
are omitted. It can be seen that the differences are more pronounced 
on the $\mathrm{B}$ state. This is not surprising, since the 
$\mathbf{K}$-$\Omega$ coupling affects transitions involving 
this state directly. Also, with increased intensity the differences 
intensify, but do not exceed 2-3\% even for the higher employed 
intensity (Fig. \ref{Fig:KER_diff} panels e) and f)). We may conclude
that, in agreement with the results obtained for the dissociation 
yield, the inclusion of electron-rotation coupling does not alter 
the KER spectra in a measurably meaningful way.

\subsection{Photofragment Angular Distribution}
\begin{figure}[!ht]
 \centering
 \includegraphics[width=0.85\textwidth]{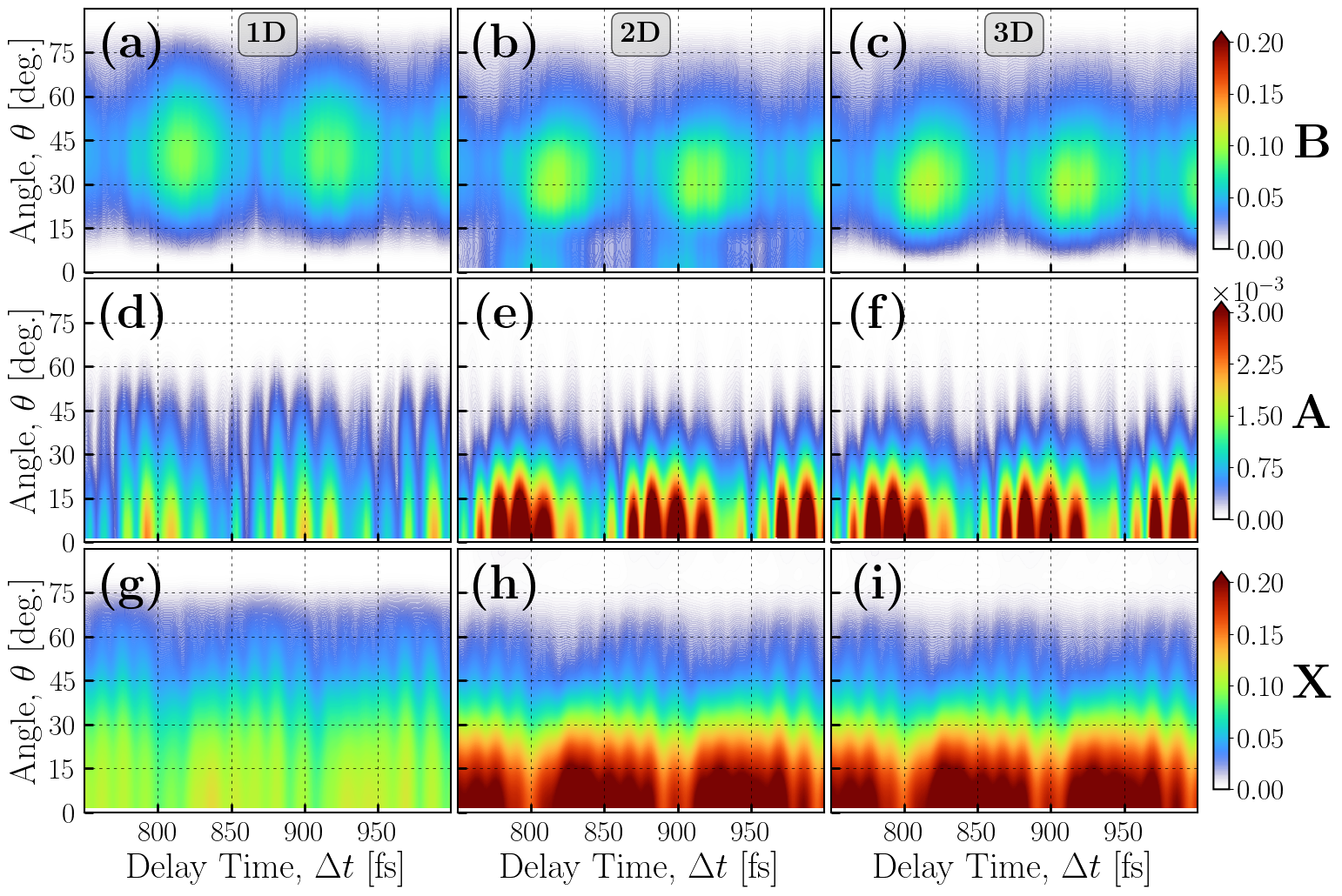}
 \caption{Photofragment angular distribution (PAD) conform 
 Eq. (\ref{Eq:PAD}) as a function of delay time for 
 $I_{pr}=1\times 10^{12}$ $\mathrm{W/cm^2}$ on each 
 dissociation channel: $\mathrm{B}$ state (a)-(c); 
 $\mathrm{A}$ state (d)-(f); $\mathrm{X}$ state (g)-(h). 
 Different columns correspond to the 1D, 2D and 3D model 
 calculations, as indicated on the upper row.}\label{Fig:PAD}
\end{figure}

Let us turn our attention to the angular distribution of the 
photofragments. Within the short delay time interval,  as we 
saw earlier for the other quantities, the PAD (Eq. \ref{Eq:PAD}) 
calculated  with the three employed models produce identical 
results. Therefore, the corresponding data is not presented 
here. Likewise, the results obtained for the two intensities 
display the same trends. Accordingly, we show in Fig. \ref{Fig:PAD} 
only the high intensity results for each electronic state and 
for the 1D, 2D and 3D models.

Two observations stand out at first glance:
i) The angular distributions on the $\Sigma$ states ($\mathrm{X}$ 
and $\mathrm{A}$) obtained with the 2D and 3D models are essentially 
indistinguishable. In contrast, within the 1D approximation -- where 
molecular rotation is completely neglected -- the results differ 
significantly from the others. The underlying mechanism is essentially 
the same as identified earlier for the interpretation of the dissociation 
yields. To be specific, due to the rotational motion induced by the 
pump pulse, the pump excited molecules start to deviate from their 
initial isotropic distribution and align with the polarization direction 
of the employed laser fields. The same symmetry of the $\mathrm{X}$ and 
$\mathrm{A}$ states means their coupling is proportional to $\cos\theta$, 
i.e. strongest along the polarization direction. As a result, a larger 
portion of the population transfers back and dissociates on the 
$\mathrm{X}$ state than in the rotationless 1D situation. The distinctive 
substructure of elongated vertical peaks is a direct consequence of the 
different vibrational components of the wave packet reaching the coupling 
region at different times.

ii) The results for the $\mathrm{B}$ state obtained with the 
2D and 3D models differ markedly in the small-angle region 
($\theta \approx 0^{\circ}-15^{\circ}$). This is a clear 
fingerprint of the inclusion of electron-rotation coupling 
into the description. Similar findings were reported by the 
group of S. B Zhang, when studying the photofragment angular 
distribution of the the $\mathrm{CH^{+}}$ and $\mathrm{MgH^{+}}$ 
molecules \cite{Zhang3, Zhang5}. Another noteworthy observation 
-- one that has not been examined previously -- is that the PAD 
data obtained with the 3D method converges to the 1D results for 
small angles. 

\begin{figure}[!ht]
 \centering
 \includegraphics[width=0.85\textwidth]{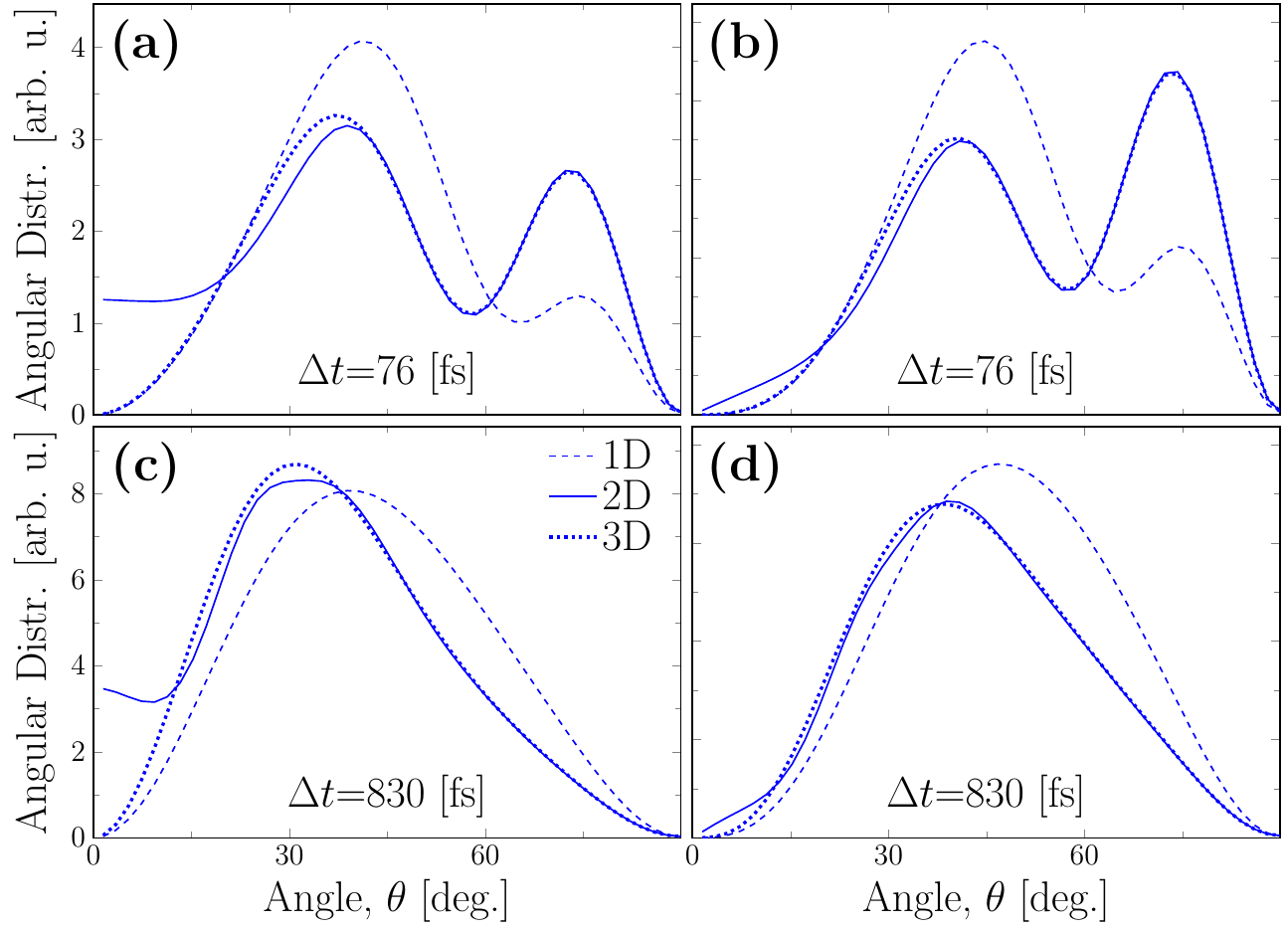}
 \caption{Angular distribution on the $\mathrm{B}$ state for 
 two delay times indicated on the figures. Panels (a) and (c) 
 show the results calculated with  Eq. (\ref{Eq:PAD}), while 
 (b) and (d) with Eq. (\ref{Eq:PAD_tot}). Dashed, continuous 
 and dotted lines represent the 1D, 2D and 3D models, respectively.}\label{Fig:PAD_cut}
\end{figure}
For easier visualization, panels a) and c) of Fig. \ref{Fig:PAD_cut}
present the angular distributions calculated with Eq. (\ref{Eq:PAD}) 
within the three models for two specific delay times and for the 
higher probe intensity. As mentioned above, the $\Sigma\leftrightarrow\Sigma$ 
transitions are proportional to $\cos\theta$, and with the chosen 
pump intensity (no Rabi floppings) the pump-excited wave packet on 
$\mathrm{A}$ has a ``$p_z$''-shape (dominantly $J=1$). On the other 
hand, $\Pi\leftarrow\Sigma$ transitions are proportional to $\sin\theta$. 
The combined effect leads in the rotationally static 1D model on 
the $\mathrm{B}$ state to a vanishig PAD both at $\theta=0^{\circ}$ 
and $90^{\circ}$, and a maxima around $\theta\sim 40^{\circ}$. When 
molecular rotation is switched on (2D, 3D), the molecules tend to 
align along the laser field, and the $p_z$ shape of the $\mathrm{A}$ 
wavepacket is skewed toward the $Z$-axis. This induces a shift of the 
$\mathrm{B}$ state PAD toward smaller angles. Moreover, the light 
induced conical intersections between the $\mathrm{A}$ and $\mathrm{B}$ 
states appear at $\theta=0^{\circ}$ (and $180^{\circ}$), which 
facilitates dissociation on $\mathrm{B}$ around, the otherwise forbidden, 
$\theta=0^{\circ}$ angles. As our data shows, incorporating also the 
$\mathbf{K}$-$\Omega$ coupling in the description entirely suppresses 
this effect, leading to no dissociating fragments along the $Z$-axis. 
For increasing laser intensity the coupling becomes stronger, i.e. 
steeper adiabatic surfaces that guide the wave packet more efficiently 
toward the LICI in 2D, and the discrepancy with the 3D model becomes 
ever higher \cite{Zhang5}.

\begin{figure}[!ht]
 \centering
 \includegraphics[width=0.85\textwidth]{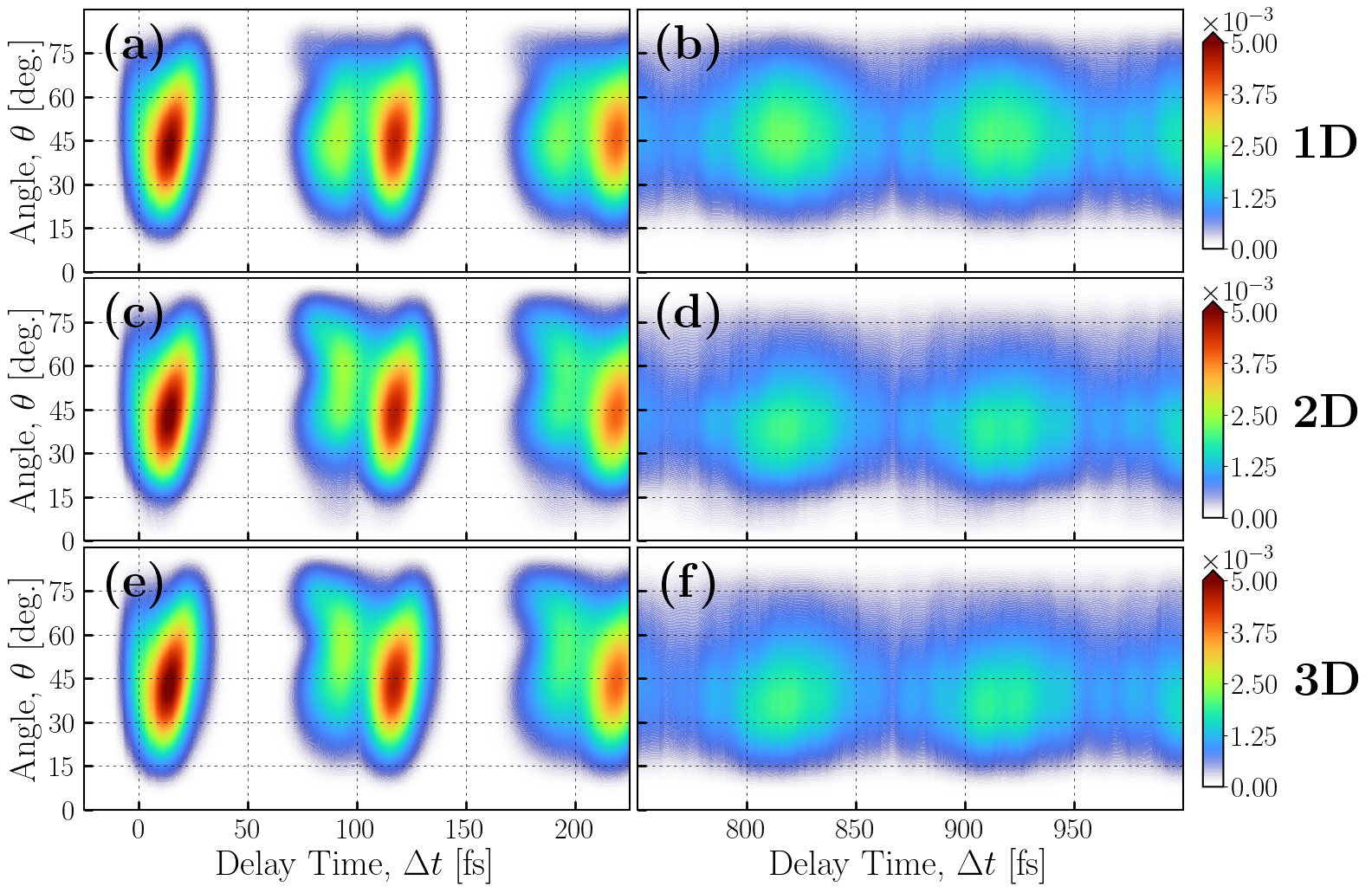}
 \caption{Total angular distribution along $\theta$ on the 
 $\mathrm{B}$ state for $I_{pr}=1\times 10^{12}$ $\mathrm{W/cm^2}$ 
 conform Eq. (\ref{Eq:PAD_tot}).}\label{Fig:PAD_scaled}
\end{figure}

Let us recall that in the dissociation probability and KER spectra 
we did not see significant differences between the 2D and 3D results. 
To better understand this discrepancy, we also calculated the total 
photofragment angular distributions $\tilde{P}_{PAD}(\theta)$ along 
the polar angle $\theta$ according to Eq. (\ref{Eq:PAD_tot}). The 
results are presented in Fig. \ref{Fig:PAD_scaled} and panels (b) 
and (d) of Fig. \ref{Fig:PAD_cut}. We can see, that the integration 
over the azimuthal angle $\varphi$ essentially washes out the 
differences between the two descriptions. This is not surprising, 
since the integration is equivalent (up to a constant factor) with 
a multiplication with the quadrature weights (i.e. $\sin\theta$), 
which diminishes the differences most strongly in the vicinity of 
$\theta=0^{\circ}$, where they are the largest.

\section{Conclusions}

In the present paper we performed pump--probe numerical simulations
to study the ultrafast molecular dynamics of the NaH molecule, considering
its first three singlet electronic states. To elucidate the underlying
mechanisms, we have investigated the combined effects of multiple
light-induced nonadiabatic couplings and nuclear rotational motion,
including the case where the electronic angular momentum projected
onto the diatomic axis is coupled to the rotational angular momentum
of the nuclei. For this reason key dynamical observables ---namely
dissociation probabilities, kinetic energy release spectra,
and photofragment angular distributions--- 
for intensities at the upper end of the linear regime have been computed. 

Based on our results, we can conclude that for dissociation probabilities
and kinetic energy release spectra, the difference between the 1D and
2D methods is much more pronounced than that between the 2D and 3D
approaches. This is understandable, since the 2D model explicitly
includes nuclear rotation and the associated light-induced nonadiabatic
effects, which are absent in the 1D treatment. In contrast, the discrepancies
between the 2D and 3D methods are comparatively minor for these observables.

However, in the case of angular distributions, we observed a substantial
difference between the 2D and 3D results in the small $\theta$ angle
region. These discrepancies intensify with increasing laser intensities. 
Interestingly, the 3D model yields results that are closer to those 
obtained with the more restrictive 1D model. 
Presently, the underlying mechanism of this behavior is not clear, 
and will be the subject of further investigations.

To gain a more accurate understanding of this behavior, we recalculated
the dissociation probabilities along the $\theta$ direction. With this
integration along the azimuthal angle $\varphi$,
the 2D and 3D models produce similar results even in the small-angle region. 
This does not imply that electron--rotation coupling has no effect; rather, 
it suggests that its impact should not be overestimated, at least for the 
considered laser intensities.


\begin{acknowledgments}
The authors are indebted to NKFIH for funding (Grant No. K146096).
The ELI ALPS project (GINOP-2.3.6-15-2015-00001) is supported by the 
European Union and co-financed by the European Regional Development Fund.
This paper was supported by the J\'{a}nos Bolyai Research Scholarship 
of the Hungarian Academy of Sciences.
\end{acknowledgments}

\section{Data Availability}
The data that support the findings of this article are openly
available \cite{data}; embargo periods may apply.

\clearpage
\appendix
\section{Laser-matter interaction in Wigner D basis}\label{Sec:Appendix1}

The operator of light-matter interaction in the dipole approximation is the scalar product 
between the $\bm{\mathcal{E}}(t)$ electric field and the dipole operator $\boldsymbol{\mu}$ (permanent or transition). 
The electric field defines the space-fixed (SF) Cartesian coordinate system, where in case of a linearly polarized laser, 
for convenience, the polarization axis coincides with the SF $Z$-axis. Dipole moments are supplied by quantum chemistry packages, 
and are defined in body-fixed (BF) Cartesian coordinates, but can be transformed to SF with the help of the rotation operator as:
\begin{align}
 \boldsymbol{\mu}^{SF} = \mathrm{\mathbf{D}}(\varphi,\theta,\chi) \boldsymbol{\mu}^{BF}
\end{align}

\begin{figure}[h!]
 \centering
 \tikzset{
    text box/.style={
        draw, thick, rounded corners=6pt,
        align=center, 
        inner xsep=0.5em},
    col1/.style={
        text width=2.3cm, minimum height=8mm},
    col2/.style={
        text width=2.8cm, minimum height=12mm},
    col3/.style={
        text width=2.3cm, minimum height=10mm},
    connector/.style={
        rounded corners=8pt, line width=2pt, 
        shorten >=2.5mm, shorten <=2.5mm,
        >={[width=7pt,length=4pt]Triangle}},
    annotate/.style={
        align=left, midway,
        inner xsep=1em, very thin},  
    subtitle/.style={
        anchor=base, inner ysep=0.5em,
        font=\bfseries\large},
    start here/.style={
        midway, above, align=center,
        font=\footnotesize, scale=0.8,
        inner sep=0.1em, yshift=2pt},
}
 \begin{tikzpicture}
  \node[text box] (eSFC) {$\mathcal{E}_{\{X,Y,Z\}}^{SF}$};
  \node[text box, below=15mm of eSFC] (eSFS) {$\mathcal{E}_{\{0,\pm 1\}}^{SF}$};
  \node[text box, right=50mm of eSFC] (uBFC) {$\mu_{\{x,y,z\}}^{BF}$};
  \node[text box, right=10mm of uBFC] (data) {Quantum\\Chemistry}; 
  \node[text box, below=15mm of uBFC] (uBFS) {$\mu_{\{0,\pm 1\}}^{BF}$};
  \node[subtitle, above=2mm of eSFC.north] (SF) {Space-fixed};
  \node[subtitle, above=2mm of uBFC.north] (SF) {Body-fixed};
  \node[text box, below=10mm of eSFS, xshift=15mm] (uSFS) {$\mu_{\{0,\pm 1\}}^{SF}$};
  \node[text box, below=30mm of eSFS] (SP) {$\bm{\mathcal{E}}\cdot\boldsymbol{\mu}=\varepsilon_{\{0,\pm 1\}}^{SF}\cdot\mu_{\{0,\pm 1\}}^{SF}$};
  
  \draw[connector, ->] (eSFC) -- (eSFS);
  \draw[connector, ->] (uBFC) -- (uBFS);
  \draw[connector, ->] (data) -- (uBFC);
  \draw[connector, ->] (uSFS) -- (SP);
  \draw[connector, ->] (uBFS.south) |- (uSFS) node[annotate, below] (WD1) {SF $\leftarrow$ BF\\with $D_{pq}^{1*}$};
  \draw[connector, ->] (eSFS) -- (SP);
  
    \coordinate (MID) at ($(eSFC.east)+(25mm,0mm)$);
    \path (current bounding box.south) coordinate[yshift=-3mm] (BOTTOM); 
    \path (current bounding box.east) coordinate[xshift=1mm] (RIGHT); 
    \path (current bounding box.north) coordinate[yshift=1mm] (TOP); 
    \path (current bounding box.west) coordinate[xshift=-2mm] (LEFT);

    \begin{scope}[on background layer]
        \fill[Green, opacity=0.8] (LEFT |- TOP) rectangle (MID |- BOTTOM);
       \fill[Salmon, opacity=0.8] (MID |- TOP) rectangle (RIGHT |- BOTTOM);
        \draw[dashed,thick] (MID |- TOP) -- (MID |- BOTTOM);
        \draw[dashed, thick] ($(LEFT)+(0,13mm)$) -- ($(RIGHT)+(0,13mm)$);
    \end{scope}

    \node[subtitle] at ($(LEFT)+(57mm,20mm)$) {Cartesian};
    \node[subtitle] at ($(LEFT)+(57mm,5mm)$) {Spherical};

  \end{tikzpicture}
\caption{Workflow for the development of laser-matter interaction.}
\end{figure}
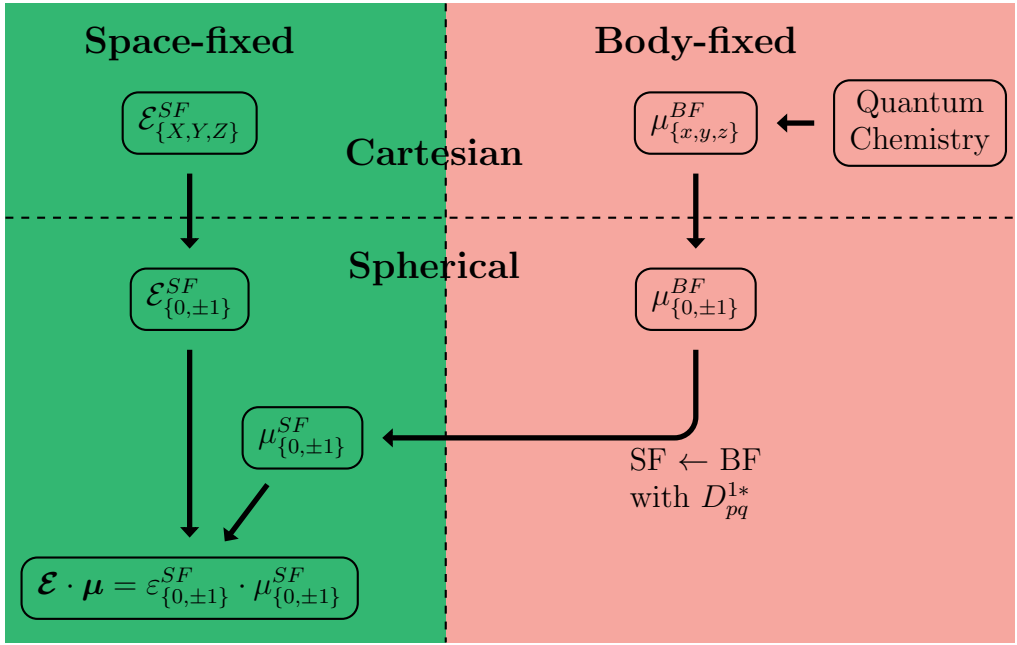

This expression is best cast in spherical coordinates, where vectors are written as 
\begin{align}
 \mathbf{v}&=\sum_{q}(-1)^{q}v_{-q}\mathbf{e}_{q}\qquad \text{with } q\in\{\pm 1, 0\}.\label{App:Eq:vec_spherical}
\end{align}
The vector components (same relations apply also for the $\mathbf{e}_{q}$ unit vectors) are related to the Cartesian ones as
\begin{align}
 v_{+1} = -\frac{1}{\sqrt{2}}\left( v_{x} +i v_{y} \right) && v_{0} = v_{z} && v_{-1} = \frac{1}{\sqrt{2}} \left( v_{x} -i v_{y} \right)\\
 v_{x}  =  \frac{1}{\sqrt{2}}\left( v_{-1} - v_{+1} \right) && v_{z} = v_{0} && v_{y} = \frac{i}{\sqrt{2}} \left( v_{-1} + v_{+1} \right).
\end{align}
Accordingly, the components of the dipole in the SF spherical coordinate system are given in terms of their BF counterparts 
\cite{Zare} as
\begin{align}
 \mu_{p}^{SF} &= \sum_{q=-1}^{1}D_{pq}^{1*}(\varphi,\theta,\chi) \mu_{q}^{BF}, \qquad \text{with }p\in\{0,\pm 1\}. \label{Eq:BF-SF}
\end{align}
Casting also the electric field in spherical form we are able to evaluate
\begin{align}
 \bm{\mathcal{E}}^{SF}\cdot\boldsymbol{\mu}^{SF} &= \sum_{p=-1}^{1} (-1)^{p}\mathcal{E}_{-p}^{SF}\mathbf{e}_{p} \cdot
 \sum_{p'=-1}^{1} (-1)^{p'}\mu_{-p'}^{SF}\mathbf{e}_{p'} =\sum_{p=-1}^{1}(-1)^{p}\mathcal{E}_{-p}^{SF}\; \mu_{p}^{SF} \nonumber\\
 %
 %
 &=\sum_{p=-1}^{1}(-1)^{p}\mathcal{E}_{-p}^{SF}\sum_{q=-1}^{1}D_{pq}^{1*}(\varphi,\theta,\chi)\mu_{q}^{BF} \nonumber\\
 &=\sum_{p=-1}^{1}\sum_{q=-1}^{1}(-1)^{p}\mathcal{E}_{-p}^{SF}\, (-1)^{q-p}D_{-p-q}^{1}(\varphi,\theta,\chi)\mu_{q}^{BF}\nonumber\\
 &=\sum_{p=-1}^{1}\sum_{q=-1}^{1}(-1)^{q}\mathcal{E}_{-p}^{SF}\,\mu_{q}^{BF}\,D^{1}_{-p\,-q}(\varphi,\theta,\chi).\label{Eq:dipole_operator_general}
 \end{align}

If one uses linearly polarized laser fields (as in the case of our pump-probe setup, 
where both pulses are linearly polarized), Eq. (\ref{Eq:dipole_operator_general})
may be simplified to a single sum, retaining the $p=0$ component for $Z$-polarized 
fields $\mathcal{E}_0 \equiv \mathcal{E}_Z$
\begin{align}
 \bm{\mathcal{E}}\cdot\boldsymbol{\mu} \equiv \mathcal{E}_{Z}^{SF} \, \mu_{Z}^{SF} &= \mathcal{E}_{Z} \sum_{q=-1}^{1} (-1)^{q}\,\mu_{q}^{BF}\, D^{1}_{0\,-q}(\varphi,\theta,\chi)\label{Eq:dipole_operator_z}\\
 &= \mathcal{E}_{Z} \left[ (-1)^{-1}\mu_{-1}^{BF}\,D_{0\,1}^{1}(\varphi,\theta,\chi) + (-1)^{0}\mu_{0}^{BF}\,D_{0\,0}^{1}(\varphi,\theta,\chi) + (-1)^{1}\mu_{1}^{BF}\,D_{0\,-1}^{1}(\varphi,\theta,\chi) \right]\nonumber\\
 %
 %
 &=  \mathcal{E}_{Z} \left[ -\frac{\mu_{x}^{BF}-i\mu_{y}^{BF}}{\sqrt{2}}\frac{\sin\theta}{\sqrt{2}}\mathrm{e}^{-i\chi} + \mu_{z}\cos\theta - \frac{\mu_{x}^{BF}+i\mu_{y}^{BF}}{\sqrt{2}}\frac{\sin\theta}{\sqrt{2}}\mathrm{e}^{i\chi} \right].\label{Eq:dipole_operator_z_explicit}
\end{align}
In the last line the explicit form of $D_{0\,q}^{1}$ was introduced
\begin{align}
 D_{0\,0}^{1}(\varphi,\theta,\chi) = \cos\theta, \qquad\text{and}\qquad D_{0\,\pm1}^{1}(\varphi,\theta,\chi) = \pm \frac{\sin\theta}{\sqrt{2}}\mathrm{e}^{\mp i\chi}.
\end{align}

Using the well known identity
\begin{align}
 \int_{0}^{2\pi}\mathrm{d}\varphi \int_{0}^{\pi}\sin\theta\,\mathrm{d}\theta \int_{0}^{2\pi}&
 \left[ D_{M_3\,\Omega_3}^{J_3}(\varphi,\theta,\chi) \right]^{*}
 D_{M_2\,\Omega_2}^{J_2}(\varphi,\theta,\chi)D_{M_1\,\Omega_1}^{J_1}(\varphi,\theta,\chi)=\nonumber\\
 &= \frac{8\pi^{2}}{2J_3 + 1} C_{J_1\, M_1\, J_2\, M_2}^{J_3\, M_3} C_{J_1\, \Omega_1\, J_2\, \Omega_2}^{J_3\, \Omega_3}\nonumber\\
 &= 8\pi^{2} (-1)^{M_3+\Omega_3}
 \begin{pmatrix}
  J_1 & J_2 &  J_3 \\
  M_1 & M_2 & -M_3
 \end{pmatrix}
 \begin{pmatrix}
  J_1 & J_2 &  J_3 \\
  \Omega_1 & \Omega_2 & -\Omega_3
 \end{pmatrix}
\end{align}
and Eq. (\ref{Eq:dipole_operator_z}) we can develop the expression 
of matrix elements in the $|JM\Omega\rangle$ basis
\begin{align}
 \left\langle J'M'\Omega' \left| \bm{\mathcal{E}} \cdot \boldsymbol{\mu} \right| JM\Omega \right\rangle &=
 \mathcal{E}_{Z}\sum_{q=-1}^{1}(-1)^{q}\,\mu_{q}^{BF}\left\langle J'M'\Omega' \left| D_{0\,-q}^{1} \right| JM\Omega \right\rangle\nonumber\\
 &\hspace*{-3cm}=\mathcal{E}_{Z}\sum_{q=-1}^{1}(-1)^{q}\,\mu_{q}^{BF}\left[ (-1)^{M'+\Omega'}\sqrt{(2J'+1)(2J+1)}
  \begin{pmatrix}
  J & 1 & J' \\
  M & 0 & -M'
 \end{pmatrix}
 \begin{pmatrix}
     J   &  1 &     J' \\
  \Omega & -q & -\Omega'
 \end{pmatrix}
 \right]\nonumber\\
 &\hspace*{-3cm}=\mathcal{E}_{Z}(-1)^{M'+\Omega'}\sqrt{(2J'+1)(2J+1)}
 \begin{pmatrix}
  J & 1 & J' \\
  M & 0 & -M'
 \end{pmatrix}
 \sum_{q=-1}^{1}(-1)^{q}\,\mu_{q}^{BF}
 \begin{pmatrix}
     J   &  1 &     J' \\
  \Omega & -q & -\Omega'
 \end{pmatrix}
 \label{Eq:Dipole_matrix_elem}
\end{align}
The Wigner 3j symbols of the last equation vanish unless the triangular 
rule of vector addition are satisfied leading to the selection rules
\begin{align}
 M' &= M, \label{Eq:M-selection}\\
 \Omega' &= \Omega-q \label{Eq:Omega-selection}.
\end{align}
The first one shows that a $Z$-polarized field leaves the projection of 
the total angular momenta unchanged. This means that the $\varphi$ Euler 
angle is redundant, and can conveniently be fixed to $\varphi=0$. 

Let us recall that all three electronic states of our model are singlet states, meaning that $\mathbf{S}=0$, therefore $\Omega\equiv\Lambda$. 
Accordingly, the selection rule (\ref{Eq:Omega-selection}) may be rewritten as $\Delta\Lambda=\Lambda'-\Lambda=q$. 
For this reason, form the expression (\ref{Eq:Dipole_matrix_elem}) of the dipole 
operator matrix elements, in case of transitions between the $\Sigma$ states 
only the $q=0-0=0$ term contributes. Note that we would get the same coupling element between $\Pi$ states, $q=1-1=0$. For 
$\Sigma \leftrightarrow \Pi$ transitions the $\Delta\Lambda_{f\leftarrow i}=\pm1$ selection rule applies. 
As a result, depending on the symmetry of the involved electronic surfaces, only one term of the light-matter coupling expression of 
Eq. (\ref{Eq:dipole_operator_z_explicit}) gives nonzero contribution, namely:
\begin{align}
 \bm{\mathcal{E}}\cdot\boldsymbol{\mu} =
 \begin{dcases}
  \quad\mathcal{E}_{0}\,\mu_{0}^{BF}\,D_{0\,0}^{1}(0,\theta,\chi) = \quad\mathcal{E}_{Z}\,\mu_{z}^{BF} \cos\theta \qquad 
  &\text{$\Sigma\to\Sigma$ (or $\Pi\to\Pi$)} \\
  -\mathcal{E}_{0}\,\mu_{+1}^{BF}\,D_{0\,-1}^{1}(0,\theta,\chi) = -\mathcal{E}_{Z}\, \frac{\mu_{x}^{BF}+i\mu_{y}^{BF}}{2}\, \sin\theta\, \mathrm{e}^{+i\chi} \qquad &\text{$\Sigma\to\Pi$}\\
  -\mathcal{E}_{0}\,\mu_{-1}^{BF}\,D_{0\, 1}^{1}(0,\theta,\chi) = -\mathcal{E}_{Z}\, \frac{\mu_{x}^{BF}-i\mu_{y}^{BF}}{2}\,\sin\theta\, \mathrm{e}^{-i\chi} \qquad &\text{$\Pi\to\Sigma$}.
 \end{dcases}
\end{align}

We also have to consider that the $\Pi$ state is doubly degenerate, with $\pi_{p_x}$ and $\pi_{p_y}$ molecular orbitals.
Both should be included in the calculations, while the resulting various physical quantities must be averaged over these 
degenerate states. This would increase the numerical demand of the calculations, since in practice it would mean four 
electronic states instead of three. To circumvent this inconvenience, we considered them as one, using the below formalism.
In case of a $\Sigma\to\Pi$ transition we can write the corresponding 
\begin{align}
 \left\langle \Pi \left| \bm{\mathcal{E}}\cdot\boldsymbol{\mu} \right| \Sigma \right\rangle 
 &= \left\langle \Pi_{p_x} \left| \bm{\mathcal{E}}\cdot\boldsymbol{\mu} \right| \Sigma \right\rangle  
  -i\left\langle \Pi_{p_y} \left| \bm{\mathcal{E}}\cdot\boldsymbol{\mu} \right| \Sigma \right\rangle \nonumber\\
 &= -\mathcal{E}_{Z}\,\frac{1}{2}\mu_{x}^{BF}\,\sin\theta\, \mathrm{e}^{i\chi} - \mathcal{E}_{Z}\, \frac{1}{2}\mu_{y}^{BF}\,\sin\theta\, \mathrm{e}^{i\chi}\nonumber\\
 &= -\mathcal{E}_{Z}\, \mu_{\perp}^{BF}\, \sin\theta\, \mathrm{e}^{i\chi}.\label{Eq:tdm_perp}
\end{align}
In the last equation we used the fact that the $x$ and $y$ components of the diplole moment are equal, and introduced the notation
\begin{align}
  \mu_{x}^{BF} = \mu_{y}^{BF} \equiv \mu_{\perp}^{BF}.
\end{align}
Transition in the opposite direction results in a matrix element that is the complex 
conjugate of (\ref{Eq:tdm_perp}), and with $\mu_{z}^{BF}\equiv\mu_{\parallel}^{BF}$ 
we arrive to our final expression
\begin{align}
 \left\langle f \left| \bm{\mathcal{E}}\cdot\boldsymbol{\mu} \right| i \right\rangle =
 \begin{dcases}
  \mathcal{E}_{Z}\, \mu_{\parallel}^{BF} \cos\theta \qquad 
  &\text{$\Sigma\to\Sigma$ (or $\Pi\to\Pi$)} \\
   -\mathcal{E}_{Z}\, \mu_{\perp}^{BF}\,\sin\theta\, \mathrm{e}^{i\chi} \qquad &\text{$\Sigma\to\Pi$}\\
   -\mathcal{E}_{Z}\, \mu_{\perp}^{BF}\, \sin\theta\, \mathrm{e}^{-i\chi} \qquad &\text{$\Pi\to\Sigma$}.
 \end{dcases}\label{Eq:tdm_final}
\end{align}

\end{document}